**Diverse genetic origins of medieval steppe nomad conquerors – a response to Mikheyev et al. (2019)**

Eran Elhaik

Lund University, Department of Biology, Lund, Sweden, 22362

* Please address all correspondence to Eran Elhaik at eran.elhaik@bio.lu.se

**Abstract**

Recently, Mikheyev et al. (2019) have produced a preprint study describing the genomes of nine Khazars archeologically dated from the 7$^{th}$ to the 9$^{th}$ centuries found in the Rostov county in modern-day Russia. Skull morphology indicated a mix of "Caucasoid" and "Mongoloid" shapes. The authors compared the samples to ancient and contemporary samples to study the genetic makeup of the Khazars and their genetic legacy and addressed the question of the relationships between the Khazar and Ashkenazic Jews. A careful examination reveals grave concerns regarding all the aspects of the study from the identification of the "Khazar" samples, the choice of environment for ancient DNA sequencing, and the analyses. The authors did not disclose the data used in their study, and their methodology is incoherent. We demonstrate that their analyses yield nonsensical results and argue that none of the claims made in this study are supported by the data unequivocally. Provided the destruction of the bone samples and the irreproducibility of the analyses, even by the forgivable standards of the field, this study is irreplicable, wasteful, and misleading.

**Introduction**

*Characterizing the samples*

The nine samples were excavated in the Rostov region. The Khazars' stronghold Sarkel was built in 833 with the help of the Byzantine emperor Theophilus to protect their north-western border. The city also served as a commercial center and controlled the Volga-Don portage, which was used by the Rus to cross from the Black Sea to the Volga and thence to the Caspian and Baltic. However, unlike Balanjar, Samandar, and Atil – Sarkel was never a political polity and thereby, burial kurgans likely belong to successful traders or military leaders who came from a broad international background and can tell us very little of the Khazar royalty. The authors' claim that the burial sites "were clearly the military and political elite of the Khaganate, which were the Khazars themselves" is not supported by the data. No military equipment was found in six out of the nine burial sites. Sample #1251 was a 40 years old hunchback man. Even the arrowheads found in the remaining three graves were different and designed to serve different purposes (Sitdikov and Karpov 2017). They could have belonged to shamans, local chieftains, or hunters rather than warriors. Most of the samples appeared in this study for the first time and are described briefly and without proper imagery, which makes it challenging to evaluate these claims. Thereby, even assuming that the samples, as handicapped as they may be, were warriors, they cannot be assumed to be Khazar soldiers. The authors have bridged the uncertainty in the identification of their samples by promoting them throughout their study: in the abstract alone the samples are termed "warriors," and then "Khazar political [elite]" and finally "ruling class."

This raises the primary question that must be first asked about Mikheyev et al.'s (2019) study – have they studies Khazar samples at all? Since the Khazars left no written evidence, their identification remains problematic and had been debated at length in the literature and any Khazarian affiliation must be fully justified. The author made several attempts to justify the assignment of Kazazian ancestry.

First, since the findings were recovered from burial kurgans, they considered them "warriors" and thereby "Khazars." However, even assuming that the nomadic graves were those of soldiers, there is no certainty that they were Khazars. It is well established that the Khazars army was an international force furnished by the Khazar elite, Muslims, Alans, Torks/Ghuzz, pagan Rus and Saqaliba or Vikings and Finns/East Slavs, and other pagan Eastern European mercenaries recruited from the most loyal subjected people (Dunlop 1967; Noonan 2001; Golden 2007). Any of those could have deemed themselves worthy or be considered worthy for a nomadic burial, much like the Egyptian mummies found to be Roman veterans (Schuenemann et al. 2017). Interestingly, some of the burial sites had Byzantine coins.

Second, the items and tools of the Saltovo-Mayatskaya culture were used by different peoples inhabiting the Khaganate. Not only this is no evidence of a Khazarian ancestry, but the nomads in the Don region during the seventh through the first half of the eighth centuries alone could only have been Khazar, Bulgar, or Onogurs. Moreover, Khazar antiquities in South Russia are not represented by Saltovo culture in any strict classical meaning of the term (Afanas'ev 2018). The relationship between the Khazars and Saltovo culture are still debated (e.g., Werbart 1996; Afanas'ev 2018) are by no means "uniting archaeological criteria that characterized Khazaria." The tradition of accompanying a human interment with a horse was so widespread in the steppe during this time that it reached Hungary and China by the ninth century. The burial of a horse was one of the characteristic features of nomad interments and did not indicate a Khazarian ancestry (Uspenskii 2018). The authors have acknowledged this difficulty.

Third, the authors have suggested that kurgans with square trenches are a hallmark of the Khazars, however clearly exceptions are allowed as no square trench was detected in the case of Sample #67. The evidence that there exists a "consensus among historians that kurgans with square trenches are specifically Khazar burial grounds" is a study by Науменко (2004), which was cited no less than five times.

Finally, ethnogenesis claims typically allude to a shared origin, genetic and geographic, shared culture, and some similarities in appearance, language, and religion. The samples in this study do not share any common characteristics. They exhibit major genetic differences in their uniparental chromosomal haplogroups as well as in their autosomal ancestry. The authors have failed to show common geographical origins, and such cannot be presumed based on the data. The allegation as if the samples belong to a minority class of Asiatic origin who mixed to various degrees over time is unconvincing and unsupported either by the data or the results. Three out of eight samples differ in their skull morphology. The two samples (#619 and #656) that exhibit the highest similarity by the admixture analysis (Figure SF3) had different skull morphologies (Table S1) and these are the earliest samples in the dataset (VII-VIII).

*Dating the samples*

Dating ancient remains is of crucial importance to produce meaningful and reliable historical reconstructions particularly considering the growing medicalization of the field. Accurate dating of the samples in the study is essential to support their identity provided the narrow timeline in which Sarkel existed as a Khazarian stronghold (833-965). By the mid-10$^{th}$ century, the fortress was captured by the Kievan Rus' and settled by the Slavs, changing the ancestry composition of the population in the region (Dunlop 1967). Provided these time constraints, it would be expected of the authors to employ radiocarbon dating, which is the gold standard to date ancient organic materials (Taylor and BarYosef 2014). However, this was not the case in this study. Instead, the authors provided a subjective date based on the archeological context. This form of interpretation has already led to misunderstandings on numerous occasions. For instance, a bone from the Darra-i-Kur cave in Afghanistan, initially assumed to be from the Palaeolithic (30,000 BP) (Dupree 1972) and often cited as one of the very few Pleistocene human fossils from Central Asia, was recently radiocarbon-dated to the Neolithic (4,500 BP) (Douka et al. 2017). Though a major limitation of radiocarbon dating is the high amount of collagen extraction (500 mg), the authors could have employed TPS (Esposito et al. 2019) to support their dating, which, excepting Sample #656, were unjustified.

*Ancient DNA sequencing environment*

Ancient DNA studies require a dedicated environment set up to prevent contamination. Because the DNA in most ancient specimens is highly degraded, the potential for contamination of ancient samples and DNA extracts with modern DNA is considerable (Fulton 2012; Knapp et al. 2012; Morozova et al.

2016). An ancient DNA facility must be separated from other lab facilities. Spatial isolation of the ancient DNA facility from the post-PCR laboratory is essential. The case of "dinosaur" DNA sequenced by Woodward et al. (1994) that had turned out to be contamination of human origin (Allard, Young, and Huyen 1995; Hedges and Schweitzer 1995) is an excellent example of why adhering to safety protocols is essential. Working in a "clean room," while satisfying the basic criteria of any lab work, is insufficient to prevent contamination and cannot guarantee reproducible and authentic results. To the best of our knowledge, the study design of Mikheyev et al. (2019) has failed to adhere to these standard requirements.

*Choice of the reference dataset*

The authors have curated modern and ancient samples from public sources to be used as a reference in their analyses. Since the choice of reference population has a direct effect on the results, it is expected that the choice of populations would be visible and justified. Regretfully, the authors have failed to report the samples which they analyze, as is the standard in population genetic studies. Some of their samples (Triska et al. 2017) are unavailable. Key populations in the Khazar scenery are missing, such as Bulgarians (Yunusbayev et al. 2011), more Turkish populations (Hodoglugil and Mahley 2012), Mongols (Pagani et al. 2016), and populations that currently inhabit the ancient lands of the Western Turkic Khaganate from where the Khazars have emerged like Tajiks (Yunusbayev et al. 2011) and Western Asian populations (Conrad et al. 2006). Key ancient people are also missing, such as the ancient Xinjiang (Ning et al. 2019) and Iron Age nomads like the Cimmerians, Scythians, and Sarmatians (Krzewińska et al. 2018) that may represent a vestige of the Khazar ancestors.

Studying the relationships between Ashkenazic Jews and the Khazars necessitates analyzing a comprehensive dataset of Ashkenazic Jews, mainly Russian and Caucasus (mountain) Jews – the likeliest progenitors of the Judaized Khazars. Remarkably, the authors do not note the source of the Ashkenazic Jews used in their analysis. The only cited dataset that includes Ashkenazic Jews (Behar et al. 2013) consists of 29 samples (13 from Western Europe and 16 from Eastern Europe), of which one is a Russian Jew and none of which are from the studied region. Surprisingly, the Caucasus Jews included in the original dataset were omitted from this study. This choice of a limited and irrelevant dataset to address a key question in the manuscript is peculiar provided the numerous datasets (e.g., Need et al. 2009; Bray et al. 2010; Das et al. 2016; Gladstein and Hammer 2016) which, unlike the so-called "Jewish HapMap Project," do not require signing a contract that limits the academic freedom of the authors. Using a small dataset is bound to produce limited results, which may lead one to conclude that "the genetic homogeneity of the worldwide Jewish population is also problematic."

A further curiosity concerns the author's emphasis on demonstrating the correlation between genetics and geography in their samples and their results for Ashkenazic Jews. Their analyses (Mikheyev et al. (2019) Figure 1A) shows that Ashkenazic Jews cluster with North Ossetia and Caucasus populations, rather than their neighboring communities. It is difficult to reconcile this finding with the authors' beliefs that the high genetic-geographic correlation makes their populations "strongly informative about the geographic origins of the Khazars" and their criticism towards past studies that showed a Caucasus origin for Ashkenazic Jews (Das et al. 2016; Das et al. 2017).

*Choice of markers*

The authors report the number of bases covered in their sequencing analysis (Table S1), but not the number of SNPs, as is expected in genomic studies. We suspect that the actual number of SNPs is very low provided the low coverage and small fraction of genome covered by more than one read. In light of that, the decision to further restrict the dataset to the AIMs reported by Elhaik et al. (2014) is puzzling. Mikheyev et al. (2019) analyzed the Q20 DP2 dataset consisting of 389 – 10049 SNPs per sample. The authors have failed to demonstrate that such a low number of SNPs in ancient genomes can produce reliable results. In our experience, these numbers for nearly haploid genomes are insufficient to derive accurate ancestry estimation (Esposito et al. 2018; Esposito et al. 2019). It is easy to show why this is the case.

Using a publicly available dataset (Lazaridis et al. 2014), we selected three ancient genomes (AG2: Upper Paleolithic hunter-gatherer from Krasnoiarsk Krai, Skoglund_farmer (formerly Gök4): Swedish Neolithic farmer; and Skoglund_HG (combined Ajv52, Ajv70, and Ire8) Swedish Neolithic hunter-gatherers) that shared 20,195, 23,579, and 22,956 SNPs, respectively, with the GenoChip dataset of ~130,000 SNPs (Elhaik et al. 2013) used by Mikheyev et al. (2019). We also included 1941 modern-day populations from the same dataset.

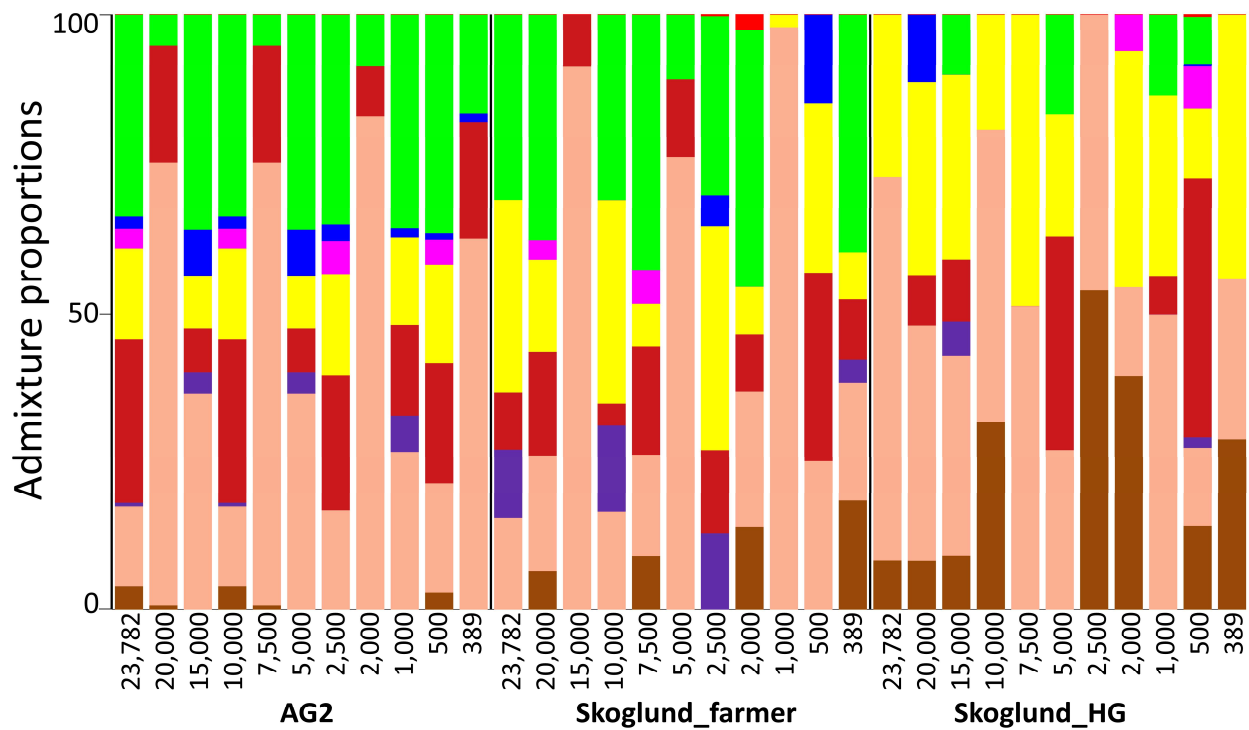

Figure 1. The admixture components of three ancient genomes calculated for a decreasing number of SNPs. Each individual is represented by a vertically stacked column of color-coded admixture components that reflects genetic contributions from nine putative ancestral populations.

Following Mikheyev et al.'s (2019) approach, we calculated the admixture components for all the samples as described in (Elhaik et al. 2014). Briefly, we employed a *supervised* admixture where the admixture components of ancient and modern populations are calculated with respect to nine gene pools. We then calculated the Euclidean distances between each ancient sample and all modern-day samples and found the sample with the shortest distance (i.e., genetically most similar). We repeated the analysis for a smaller number of SNP sets. Figure 1 illustrates the instability of the admixture components when calculated over a decreasing number of SNPs. Unsurprisingly, this has a direct effect on the calculations of the nearest population (Table 1) and its geolocation (Figure 2).

Table 1. The most similar modern-day population to each ancient genome, inferred on SNP sets that decrease in size.

| #SNP | Ancient DNA sample | Closest modern-day sample |
|---:|---|---|
| 23,782 | AG2 | Siberia_Aleut |
| 20,000 | AG2 | Siberia_Aleut |
| 15,000 | AG2 | Russia |
| 10,000 | AG2 | Siberia_Aleut |
| 7,500 | AG2 | Siberia_Aleut |
| 5,000 | AG2 | Russia |
| 2,500 | AG2 | Russia |
| 2,000 | AG2 | Russia |
| 1,000 | AG2 | Siberia_Aleut |
| 500 | AG2 | Turkey_Adana |
| 389 | AG2 | Turkey_Istanbul |
| 23,782 | Skoglund_farmer | Israel_Bedouin |
| 20,000 | Skoglund_farmer | Israel_Bedouin |
| 15,000 | Skoglund_farmer | Yemen_Jew |
| 10,000 | Skoglund_farmer | Italy_Sardinia |
| 7,500 | Skoglund_farmer | Yemen_Jew |
| 5,000 | Skoglund_farmer | Italy_Sardinia |
| 2,500 | Skoglund_farmer | Israel_Bedouin |
| 2,000 | Skoglund_farmer | Italy_Sardinia |
| 1,000 | Skoglund_farmer | Russia_Nogai |
| 500 | Skoglund_farmer | Russia_Yukagir_Forest |
| 389 | Skoglund_farmer | Pakistan_Burusho |
| 23,782 | Skoglund_HG | Spain_Pais_Vasco_IBS |
| 20,000 | Skoglund_HG | Spain_Pais_Vasco_IBS |
| 15,000 | Skoglund_HG | Russia |
| 10,000 | Skoglund_HG | Russia_Tlingit |
| 7,500 | Skoglund_HG | Siberia_Aleut |
| 5,000 | Skoglund_HG | Russia_Tlingit |
| 2,500 | Skoglund_HG | Indian_Gujarati_GIH |
| 2,000 | Skoglund_HG | Russia_Nogai |

| 1,000 | Skoglund_HG | Bolivian_LaPaz |
| 500   | Skoglund_HG | Russia_Tlingit |
| 389   | Skoglund_HG | Eskimo_Naukan  |

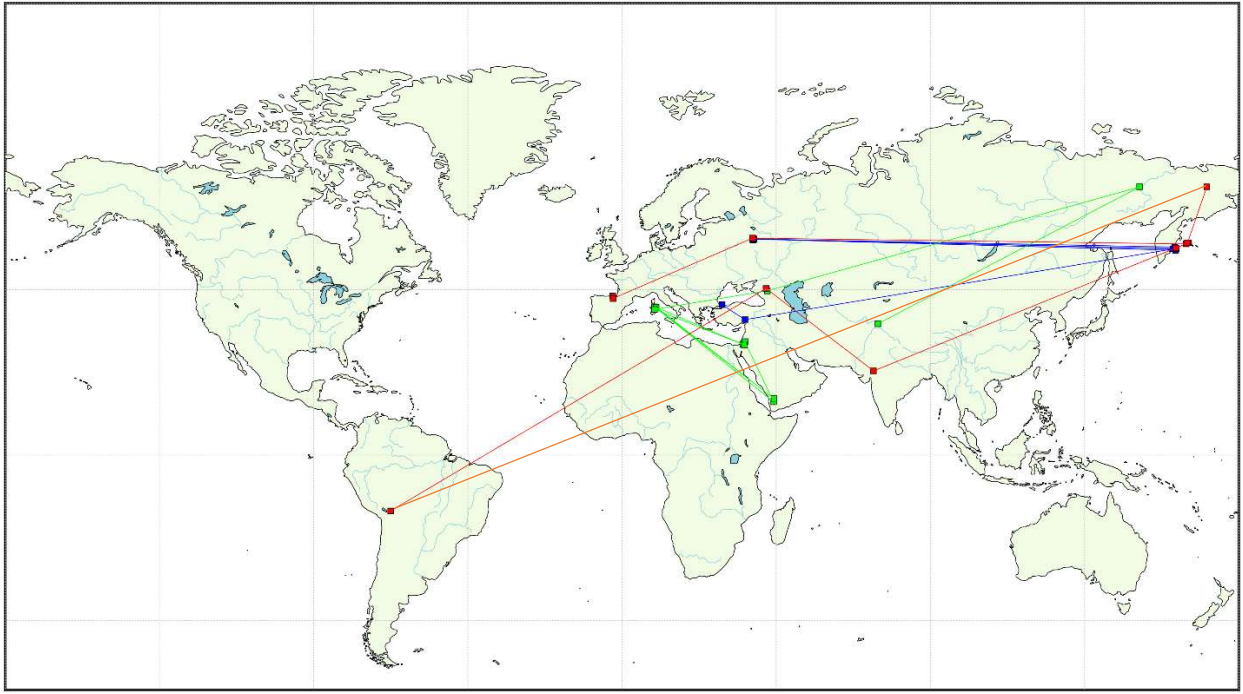

Figure 2. The geographical location of the nearest populations (squares) found for each ancient genome, color-coded in blue (AG2), green (Skoglund_farmer), and red (Skoglund_HG). Color-coded lines connect consecutive predictions.

Since the nearest populations vary across at least two continents, it is clear that no conclusion can be drawn from these analyses. We note that in our study all the ancient genomes shared the same SNPs, whereas in Mikheyev et al. (2019) the samples had different SNPs that may not exhibit a broad geographical dispersion which renders their results even more spurious.

*Genomic analyses*

The authors carried out four autosomal analyses and one uniparental chromosomal analysis where they aimed to "explore the ethnic composition of the Khazars, and also specifically test the hypothesis of their relatedness to contemporary Ashkenazi Jews." However, contrary to what has been claimed, the authors did not perform the analyses necessary to test "relatedness" but rather genetic similarity.

1. *MDS (autosomal)*

Multidimensional scaling (MDS) is a dimension reduction technique that provides a *k*-dimensional representation of any substructure, similarly to principal components analysis (PCA) (Purcell et al. 2007). The authors applied MDS to modern-day samples assembled from the literature. However, they avoid explaining how they calculated the MDS. The purpose of this analysis, applied to modern populations, was to show that their genetic similarity resembles their geographic location, which was estimated qualitatively and can be debatable as to what extent the joint clustering of French and Hungarians geographically separated from each other by over 1500km can be considered "similar." The authors next claim that due to the genetic-geographic similarity between these modern-day populations they are "strongly informative about the geographic origins of the Khazars" who lived over 1,000 years ago. It is hard to see the logic in such an argument, particularly when ancient DNA data from the relevant regions are readily available and can be used to study the Khazar origins more thoroughly.

2. *GPS and reAdmix (autosomal)*

The authors note that "The GPS algorithm determines a single point on the map that corresponds to the location of individuals with the most similar genotype" but refrain from showing the results. Instead, they used GPS and reADmix in a similar manner with the aim of inferring the genetic similarity between the ancient and modern-day genomes by comparing the nine admixture components and finding the most similar genomes. We have shown above (Figures 1-2) why such analyses yield nonsensical results and why no reliable conclusions can be drawn from these analyses.

3. *Measuring the distances to other ancient genomes (autosomal)*

The authors claim to incorporate "587 ancient Eurasians (1 BCE - 3000 BCE) to the GPS/reAdmix database" to understand the potential relationships between the Khazars and Bronze Age populations. Their report that most of the Khazar samples are related is evidence in Figure SF4. Since it is unclear how this figure was calculated, it cannot be replicated. In this figure, six out of nine Khazar samples cluster together and appear as an outgroup to the remaining 29 ancient genomes. There are several reasons to explain this clustering. First, that the poor quality of the "Khazarian" genomes and their small number of SNPs rendered them dissimilar to all the other samples; and second, that the most different 29 ancient genomes remained after the omission of 95% of the initial dataset (558/587). Unfortunately, a careful selection of data to produce the desired results is a widely used approach in population genetic studies and has been criticized before (Elhaik 2016; Marshall et al. 2016).

4. *"TreeMix F3 outgroup analysis" (autosomal)*

The last analysis is a three-population test (Reich et al. 2009). This test is of the form f3(A; B, C), where a significantly negative value of the f3 statistic implies that population *A* had a complex history. The authors do not note which populations were used as *B* and how many SNPs the tested samples shared, which would have a direct effect on the test results. It is even less clear why a modern-day population living in the Philippines should be considered the outgroup for the ancient "Khazar" samples or even why an admixture test is necessary after the authors employed two admixture-based tests.

5. *Haplogroup analysis (mtDNA and Y chromosomes)*

The authors identified the mitochondrial and Y chromosomal haplogroups. They compared these haplogroups with a reference population database of unknown size and content and reported "no significant trace of Ashkenazi genetic composition in either nuclear, mitochondrial or Y-chromosome data" which is "strongly indicating that Khazars were generally not related to them."

This report is inconsistent with the literature. Analyzing 367 Ashkenazic Jews, Das et al. (2016) reported that the H1 Subclade is common among Ashkenazic Jews (11%) with Subclades H5 (1.1%) being rarer. One sample with the maternal haplogroup H1a3 (as Sample #619) had a Siberian paternal haplogroup of Q1b1a and one sample with maternal haplogroup X2e (as Sample #531) had a Caucasus paternal haplogroup of G2a1. Concerning the Y chromosomal haplogroup, 2.8% Ashkenazic Jewish males belong to haplogroup Q (as Sample #619).

Several points are noteworthy from this analysis: first, that similarity in haplogroups cannot support or disprove the relationships between Ashkenazic Jews and the Khazars, provided the great heterogeneity of the two groups. The apparent heterogeneity of the "Khazars" apparent in all the analyses casts doubts on the ability to consider them a single group; second that Das et al.'s (2016) study is sufficient to refute the authors' allegation of the "genetic homogeneity of the worldwide Jewish population," employed to excuse their small sample sizes; and third that the authors hypothesis that Ashkenazic Jews have descended from a minority group of Khazar royalty is a novel absurdity conjectured by Mikheyev et al. (2019). Even assuming that the Ashkenazic Jews who descended directly from the Khazar royalty survived to modern time, a comprehensive analysis of various Ashkenazic Jews would have still yielded "no significant trace of Ashkenazi genetic composition in either nuclear, mitochondrial or Y-chromosome data."

**Discussion**

Following Klyosov and Faleeva (2017) who sequenced the first Y chromosomes of the Khazar, Mikheyev et al. (2019) have pioneered the first ancient genome study of nine Khazar samples deemed to belong to upper (warrior) class. We have shown that consistently throughout their study Mikheyev et al. (2019) carried out experiments and analyses that are inconsistent with the standards in Paleogenomics and population genetics. We questioned the authors' choice of analyses, populations, and markers and demonstrated that the authors' analyses are biased and lead to erroneous conclusions.

Population geneticists are expected to familiarize themselves with the historical and genetic literature of the populations they aim to study, in this case, the literature that discusses the Judaization of the Khazar and their relationships with Ashkenazic Jewry as well as past efforts to sequence Khazar bones (Klyosov and Faleeva 2017), some of which were used in the current manuscript, to avoid purporting absurd hypotheses and plagiarizing studies. Briefly, the Khazarian hypothesis (Koestler 1976; Sand 2009) states that following their Judaization, many of the subjugated people of the Khazars converted to Judaism and mixed with the Iranian and Byzantine Jews who migrated to Khazaria. Following its destruction, these people migrated to Eastern and Northern Europe and bolstered the Ashkenazic Jewry communities. Das et al. (2016) showed that Ashkenazic Jews trace their origins to "Ancient Ashkenazic," South to the Black Sea. From there, Jews could have penetrated Europe from two directions: through Western Turkey or by crossing Khazaria and entering from Eastern Europe after the demise of the Empire. This hypothesis differs from the Khazarian hypothesis in predating the conversion event from the 8$^{th}$ century to the early centuries AD and in identifying the God-fearers and other Iranians, Greco-Romans, and Slavs as the progenitors of Ashkenazic Jews. For this reason, Ashkenazic Jews exhibit high genetic similarity to Caucasus populations, as shown in Figure 1A in Mikheyev et al. (2019). The similarity between Ashkenazic Jews and the Khazar elite would be limited to those who retained their Khazarian genetic print.

We too share the excitement in sequencing ancient populations to increase our historical knowledge and understand their contribution to modern-day populations. However, care must be taken in justifying the methodology, demonstrating its accuracy on test datasets, and producing replicable analyses to ensure the stringency of the results. Unfortunately, this was not done in this case and may not be done in the future, due to the destruction of the samples, leaving the origin and genetic legacy of the Khazars an unresolved mystery.

**Competing interests**

EE is consultant to DNA Diagnostic Centre and DNA Consultants.